\documentclass[aps,prd,twocolumn,tightenlines,superscriptaddress,showpacs,amsmath,amssymb]{revtex4}
\usepackage{graphicx} % Include figure files
\usepackage{epstopdf} % Allow to include eps files
\usepackage{dcolumn} % Align table columns on decimal point
\usepackage{xcolor} % Color support
\usepackage{amsmath,amssymb} % Extended symbol collection
\usepackage{hyperref} % Generate pdfs with hyperlinks
\usepackage[utf8]{inputenc} % Allow UTF8 characters
\usepackage[english]{babel} % All publications are in English
\usepackage[mathlines]{lineno} % Line numbers for drafts
\usepackage{blindtext} % For testing
\usepackage{ifthen} % for conditional statements
\usepackage{orcidlink} % for ORCIDs in author list

\usepackage[noabbrev,capitalise]{cleveref} % for easier references

\usepackage{afterpage} % to clearpage of floats without a visible page break
\usepackage{rotating,booktabs,multirow} % for nicer tables

\graphicspath{{figures/}} % Use figures directory for figures

\newboolean{articletitles}
\setboolean{articletitles}{true} % False removes titles in references

\newcommand*\patchAmsMathEnvironmentForLineno[1]{%
\expandafter\let\csname old#1\expandafter\endcsname\csname #1\endcsname
\expandafter\let\csname oldend#1\expandafter\endcsname\csname
end#1\endcsname
 \renewenvironment{#1}%
   {\linenomath\csname old#1\endcsname}%
   {\csname oldend#1\endcsname\endlinenomath}%
}
\newcommand*\patchBothAmsMathEnvironmentsForLineno[1]{%
  \patchAmsMathEnvironmentForLineno{#1}%
  \patchAmsMathEnvironmentForLineno{#1*}%
}
\AtBeginDocument{%
\patchBothAmsMathEnvironmentsForLineno{equation}%
\patchBothAmsMathEnvironmentsForLineno{align}%
\patchBothAmsMathEnvironmentsForLineno{flalign}%
\patchBothAmsMathEnvironmentsForLineno{alignat}%
\patchBothAmsMathEnvironmentsForLineno{gather}%
\patchBothAmsMathEnvironmentsForLineno{multline}%
\patchBothAmsMathEnvironmentsForLineno{eqnarray}%
}

\input{belle2-symbols}

\newcommand{\BelleLumi}{980\invfb}
\newcommand{\BelleIILumi}{428\invfb}

\newcommand{\Acp}{\ensuremath{A_{\CP}}\xspace}
\newcommand{\Araw}{\ensuremath{A_{\rm raw}}\xspace}

\newcommand{\mDzpi}{\ensuremath{m(\Dz\pip)}\xspace}
\newcommand{\mD}{\ensuremath{m(\KS\KS)}\xspace}
\newcommand{\mKK}{\ensuremath{m(\Kp\Km)}\xspace}
\newcommand{\Lminsig}{\ensuremath{S_\text{min}(\KS)}\xspace}

\newcommand{\DzToKSKS}{\ensuremath{\Dz\to\KS\KS}\xspace}
\newcommand{\DzbToKSKS}{\ensuremath{\Dzb\to\KS\KS}\xspace}
\newcommand{\DzToKpKm}{\ensuremath{\Dz\to\Kp\Km}\xspace}
\newcommand{\DzbToKpKm}{\ensuremath{\Dzb\to\Kp\Km}\xspace}
\newcommand{\DzToKSpipi}{\ensuremath{\Dz\to\KS\pip\pim}\xspace}

\newcommand{\KSKS}{\ensuremath{\KS\KS}\xspace}
\newcommand{\KK}{\ensuremath{\Kp\Km}\xspace}

\newcommand{\pdf}{\ensuremath{P}\xspace}
\newcommand{\bkg}{\ensuremath{\text{non-peak}}\xspace}

\newcommand{\resBelle}{-1.1}
\newcommand{\statBelle}{1.6}
\newcommand{\systBelle}{0.1}

\newcommand{\resBelleII}{-2.2}
\newcommand{\statBelleII}{2.3}
\newcommand{\systBelleII}{0.1}

\newcommand{\resComb}{-1.4}
\newcommand{\statComb}{1.3}
\newcommand{\systComb}{0.1}

\begin{document}
\includegraphics[width=3cm]{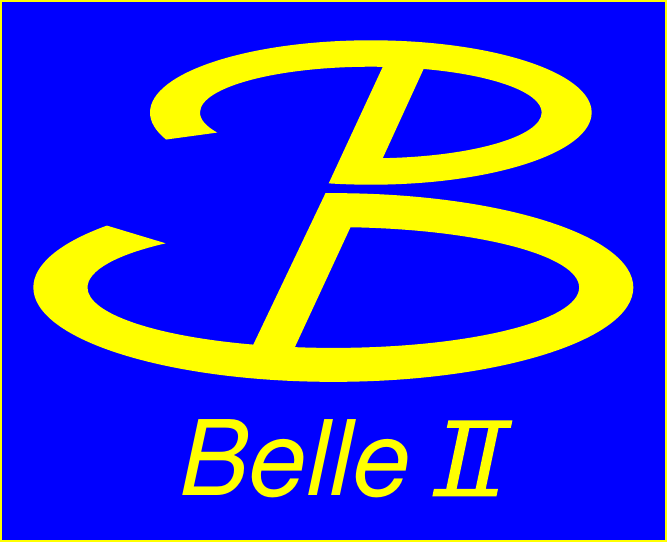}\vspace*{-1.9cm}

\begin{flushright}
Belle II Preprint 2024-026\\
KEK Preprint 2024-26
\end{flushright}\vspace{1.5cm}

\title{% WRITE THE TITLE IN THIS FILE.
{\LARGE\bfseries\boldmath Measurement of the time-integrated \CP asymmetry in $\Dz\to\KS\KS$ decays using Belle and Belle II data}
}
%%% Paper:    D0 to KS KS
%%% Journal:  Physical Review D
%%% Contacts: S. Bharati Das, K. Lalwani, A. Di Canto
%%% ====================================================================
%%% Use \input{pub067-orcid} to insert this material into your latex file.
  \author{I.~Adachi\,\orcidlink{0000-0003-2287-0173}} % 2590
% \author{K.~Adamczyk\,\orcidlink{0000-0001-6208-0876}} % 2239
  \author{L.~Aggarwal\,\orcidlink{0000-0002-0909-7537}} % 10083
% \author{P.~Ahlburg\,\orcidlink{0000-0002-9832-7604}} % 2367
  \author{H.~Ahmed\,\orcidlink{0000-0003-3976-7498}} % 11323
% \author{J.~K.~Ahn\,\orcidlink{0000-0002-5795-2243}} % 7423
  \author{H.~Aihara\,\orcidlink{0000-0002-1907-5964}} % 2223
  \author{N.~Akopov\,\orcidlink{0000-0002-4425-2096}} % 9443
  \author{A.~Aloisio\,\orcidlink{0000-0002-3883-6693}} % 2194
% \author{S.~Al~Said\,\orcidlink{0000-0002-4895-3869}} % 6823
  \author{N.~Althubiti\,\orcidlink{0000-0003-1513-0409}} % 21524
% \author{L.~Andricek\,\orcidlink{0000-0003-1755-4475}} % 2098
% \author{M.~Angelsmark\,\orcidlink{0000-0003-4745-1020}} % 13963
  \author{N.~Anh~Ky\,\orcidlink{0000-0003-0471-197X}} % 2218
  \author{D.~M.~Asner\,\orcidlink{0000-0002-1586-5790}} % 4684
  \author{H.~Atmacan\,\orcidlink{0000-0003-2435-501X}} % 2538
% \author{V.~Aulchenko\,\orcidlink{0000-0002-5394-4406}} % 8183
% \author{T.~Aushev\,\orcidlink{0000-0002-6347-7055}} % 3747
% ...
% 13403
% \author{S.~Zou\,\orcidlink{0000-0003-3377-7222}} % 19363
\collaboration{The Belle and Belle II Collaborations}

\begin{abstract}
% WRITE THE ABSTRACT IN THIS FILE.
We measure the time-integrated \CP asymmetry in \DzToKSKS decays reconstructed in $e^+e^-\to\ccbar$ events collected by the Belle and Belle II experiments. The corresponding data samples have integrated luminosities of \BelleLumi and \BelleIILumi, respectively. The \Dz decays are required to originate from the $\Dstarp\to\Dz\pip$ decay, which determines the charm flavor at production time. A control sample of \DzToKpKm decays is used to correct for production and detection asymmetries. The result, $(\resComb\pm\statComb\stat\pm\systComb\syst)\%$, is consistent with previous determinations and with \CP symmetry.
\end{abstract}

\maketitle

% WRITE THE MAIN TEXT IN THIS FILE.
\section{Introduction}
Charge-parity (\CP) violation in the charm sector was first observed by the LHCb collaboration in the difference between the \CP asymmetries of $\Dz\to\Kp\Km$ and $\Dz\to\pip\pim$ decays~\cite{Aaij:2019kcg}. (Throughout this paper, charge-conjugate modes are implied unless stated otherwise.) A more recent result reports evidence that the \CP violation occurs primarily in the $\Dz\to\pip\pim$ mode~\cite{LHCb:2022lry}. However, the origin of the LHCb \CP asymmetry is not fully understood and there is a debate about whether it could be due to physics beyond the standard model~\cite{Chala:2019fdb,Grossman:2019xcj,Schacht:2021jaz,Schacht:2022kuj}. This motivates additional measurements of \CP asymmetries in $D$ decays to two pseudoscalars to evaluate predictions based on the pattern of flavor-$SU(3)$ breaking in charm decays~\cite{Grossman:2019xcj,Buccella:2019kpn}. 

The \DzToKSKS decay proceeds through a color- and Cabibbo-suppressed transition that involves interference between $c\to us\bar{s}$ and $c\to ud\bar{d}$ amplitudes, mediated by the exchange of a $W$ boson at tree level. Such interference can generate \CP asymmetries at the 1\% level, even if the Cabibbo-Kobayashi-Maskawa phase is the only source of \CP violation~\cite{Buccella:2019kpn,Brod:2011re,Nierste:2015zra}. These features make the \DzToKSKS mode an important ingredient in understanding of the origin of \CP violation in charm decays.

Several experiments have searched for \CP violation in \DzToKSKS decays~\cite{CLEO:2000opx,LHCb:2015ope,Dash:2017heu,LHCb:2021rdn,CMS:2024hsv}. The world-average value of the \CP asymmetry, $\Acp(\DzToKSKS)=(-1.9\pm1.0)\%$~\cite{HFLAV:2022pwe}, is dominated by measurements from Belle~\cite{Dash:2017heu} and LHCb~\cite{LHCb:2021rdn}. Using a sample of \epem collisions with an integrated luminosity of 921\invfb, Belle measured $\Acp(\DzToKSKS)=(-0.02 \pm 1.53 \pm 0.02 \pm 0.17)\%$, where the first uncertainty is statistical, the second systematic, and the third is due to the uncertainty in the \CP asymmetry of the $\Dz\to\KS\piz$ reference mode. A more precise result is obtained by LHCb using a 6\invfb sample of $pp$ collisions: $(-3.1\pm1.2\pm0.4\pm0.2)\%$, where the first uncertainty is statistical, the second is systematic, and the third is due to the uncertainty in the \CP asymmetry of the reference \DzToKpKm decay. The measurement of $\Acp(\Dz\to\Kp\Km)$ has been recently improved by LHCb~\cite{LHCb:2022lry}, bringing the corresponding uncertainty below the $0.1\%$ level. The precision of the world-average value of $\Acp(\DzToKSKS)$ is therefore limited by statistical uncertainties and is expected to greatly improve over the next few years with the larger samples being collected at LHCb and Belle II.

In this paper, we report a measurement of $\Acp(\DzToKSKS)$ using a combination of Belle and Belle II data, which have integrated luminosities of \BelleLumi and \BelleIILumi~\cite{lumi}, respectively. To determine the production flavor of the neutral \D meson, which is referred to as \textit{tagging}, we reconstruct \DzToKSKS decays from $\Dstarp\to\Dz\pip$ decays in $e^{+}e^{-}\to \ccbar$ events. The decay-time-integrated \CP asymmetry, defined as
\begin{equation}
\Acp(\DzToKSKS) = \frac{\Gamma(\DzToKSKS)-\Gamma(\DzbToKSKS)}{\Gamma(\DzToKSKS)+\Gamma(\DzbToKSKS)}\,,
\end{equation}
where $\Gamma$ is the time-integrated decay rate, is measured from the \textit{raw} asymmetry in reconstructed yields,
\begin{equation}
\Araw^{\KSKS} = \frac{N(\DzToKSKS)-N(\DzbToKSKS)}{N(\DzToKSKS)+N(\DzbToKSKS)}\,,
\end{equation}
after correcting for \textit{nuisance} asymmetries induced by production and detection mechanisms. The nuisance asymmetries are determined using the control channel $\Dstarp\to\Dz(\to\KK)\pip$. In the limit of small \CP and nuisance asymmetries, the raw asymmetry of a \Dstarp-tagged \Dz decay to a two-body and \CP-symmetric final state $f$ can be expressed as
\begin{equation}
\Araw^f = \Acp(\Dz\to f) + A_{\rm P}^{\Dstarp}(\Dz\to f) + A^{\pi}_{\epsilon}(\Dz\to f)\,.
\end{equation}
The term $A_{\rm P}^{\Dstarp}$ arises, in $\epem\to\ccbar$ events, from the forward-backward asymmetric production of charm hadrons due to $\gamma^{*}$-$Z^{0}$ interference and higher-order QED effects~\cite{Berends:1973fd,Brown:1973ji,Cashmore:1985vp}. This production asymmetry is an odd function of the cosine of the polar angle in the center of mass system, $\cos\theta^*(\Dstarp)$, and therefore averages to zero only when integrated over a $\cos\theta^*(\Dstarp)$-symmetric acceptance, which is not the case at Belle and Belle II. The term $A^{\pi}_{\epsilon}$ is the detection asymmetry of the low-momentum tagging pion from the \Dstarp decay (\textit{soft} pion). Having opposite strangeness, the two kaons in the \Dz final states do not contribute any \CP violation or detection asymmetry. Assuming that the nuisance asymmetries are identical for \DzToKSKS and \DzToKpKm decays, or that they can be made so by weighting the control sample, the \CP asymmetry in \DzToKSKS decays can be determined as 
\begin{multline}
\Acp(\DzToKSKS) = \Araw^{\KSKS} - \Araw^{\KK}\\ + \Acp(\DzToKpKm)\,,
\end{multline}
where $\Acp(\DzToKpKm)$ is an external input. In particular, we use $\Acp(\DzToKpKm)$ computed from the results reported in Refs.~\cite{LHCb:2022lry,LHCb:2021vmn} as
\begin{align}\label{eq:inputAcpD0KK}
\Acp(\DzToKpKm) &= \Acp^{\rm dir}(\DzToKpKm) + \Delta Y \nonumber\\&= (6.7\pm5.4)\times10^{-4}\,,
\end{align}
where $\Acp^{\rm dir}$ is the direct \CP asymmetry, $\Delta Y$ is the asymmetry arising from \CP violation in mixing and in the interference between mixing and decay~\cite{Grossman:2006jg,CDF:2011ejf}, and we have used the fact that the average decay time of the selected $\Dz\to\Kp\Km$ decays is equal to the \Dz lifetime. The value of $\Delta Y$ is here assumed to be independent of the \Dz decay mode~\cite{Grossman:2006jg}. The measurements of $\Acp^{\rm dir}(\DzToKpKm)$ from Ref.~\cite{LHCb:2022lry}, $(7.7\pm5.7)\times10^{-4}$, and $\Delta Y$ from Ref.~\cite{LHCb:2021vmn}, $(-1.0\pm1.1)\times10^{-4}$, have a total correlation of 35\%~\cite{Serena}.

To avoid potential bias, an arbitrary and undisclosed offset is added to the measured value of $\Araw^{\KSKS}$ from the data. This offset remains undisclosed until we finalize the entire analysis procedure and determine all uncertainties.

The paper is organized as follows. \Cref{sec:detector} provides an overview of the Belle and Belle II detectors. \cref{sec:simulation} details the simulation samples used in the measurement. The reconstruction and selection of both the signal \DzToKSKS and control \DzToKpKm decays are presented in \cref{sec:selection}. \Cref{sec:equalization} discusses the weighting of the control mode to match the kinematic distributions of the signal mode and ensure an accurate cancellation of the nuisance asymmetries. Determination of the raw asymmetries is covered in \Cref{sec:fit}, followed by a discussion of the systematic uncertainties affecting the measurement in \cref{sec:systematics}. Final results are presented in \cref{sec:results}, followed by concluding remarks.

\section{Belle and Belle II detectors\label{sec:detector}}
The Belle experiment~\cite{belle_detector,Brodzicka:2012jm} operated at the KEKB asymmetric-energy $\epem$ collider~\cite{kekb,Abe:2013kxa} between 1999 and 2010. The detector consisted of a large-solid-angle spectrometer, which included a double-sided silicon-strip vertex detector, a 50-layer central drift chamber, an array of aerogel threshold Cherenkov counters, a barrel-like arrangement of time-of-flight scintillation counters, and an electromagnetic calorimeter composed of CsI(Tl) crystals. All subdetectors were located inside a superconducting solenoid coil that provided a 1.5 T magnetic field. An iron flux-return yoke, placed outside the coil, was instrumented with resistive-plate chambers to detect \KL mesons and identify muons.  Two inner detector configurations were used: a 2.0\cm radius beam pipe and a three-layer silicon vertex detector; and, from October 2003, a 1.5\cm radius beam pipe, a four-layer silicon vertex detector, and a small-inner-cell drift chamber~\cite{Natkaniec:2006rv}.

The Belle II detector~\cite{b2tech,Kou:2018nap} is an upgrade with several new subdetectors designed to handle the significantly larger beam-related backgrounds of the new collider, SuperKEKB~\cite{Akai:2018mbz}. It consists of a silicon vertex detector wrapped around a 1\cm radius beam pipe and comprising two inner layers of pixel detectors and four outer layers of double-sided strip detectors, a 56-layer central drift chamber, a time-of-propagation detector, an aerogel ring-imaging Cherenkov detector, and an electromagnetic calorimeter, all located inside the same solenoid as used for Belle. A flux return outside the solenoid is instrumented with resistive-plate chambers, plastic scintillator modules, and an upgraded read-out system to detect muons and \KL mesons. For the data used in this paper, collected between 2019 and 2022, only part of the second layer of the pixel detector, covering 15\% of the azimuthal angle, was installed.

\section{Simulation\label{sec:simulation}}
We use simulated event samples to identify sources of background, optimize selection criteria, match the kinematic distributions of signal and control decays, determine fit models, and validate the analysis procedure. We generate $\epem\to\Upsilon(nS)$ ($n=4,5$) events and simulate particle decays with \textsc{EvtGen}~\cite{Lange:2001uf}; we generate continuum $\epem\to\qqbar$ (where $q$ is a $u$, $d$, $c$, or $s$ quark) with \textsc{Pythia6}~\cite{Sjostrand:2006za} for Belle, and with \textsc{KKMC}~\cite{Jadach:1990mz} and \textsc{Pythia8}~\cite{Sjostrand:2014zea} for Belle~II; we simulate final-state radiation with \textsc{Photos}~\cite{Barberio:1990ms,Barberio:1993qi}; we simulate detector response using \textsc{Geant3}~\cite{Brun:1073159} for Belle and \textsc{Geant4}~\cite{Agostinelli:2002hh} for Belle II. In the Belle simulation, beam backgrounds are taken into account by overlaying random trigger data. In the Belle II simulation, they are accounted for by simulating the Touschek effect~\cite{PhysRevLett.10.407}, beam-gas scattering, and luminosity-dependent backgrounds from Bhabha scattering and two-photon QED processes~\cite{Lewis:2018ayu,Natochii:2023thp}.

\section{Reconstruction and event selection\label{sec:selection}}
We use the Belle II analysis software framework (basf2) to reconstruct both Belle and Belle~II data~\cite{Kuhr:2018lps,basf2-zenodo}. The Belle data are converted to the Belle II format for basf2 compatibility using the B2BII framework~\cite{Gelb:2018agf}. 

Events are selected by a trigger based on either the total energy deposited in the electromagnetic calorimeter or the number of charged-particle tracks reconstructed in the central drift chamber. The efficiency of the trigger is found to be close to 100\% for both signal and control mode decays.

Candidate $\KS\to\pip\pim$ decays are reconstructed from combinations of oppositely charged particles that are constrained to originate from a common vertex. These particles are assumed to be pions and the resulting dipion mass is required to be in the $[0.45,0.55]\gevcc$ range. Pairs of \KS candidates are combined to form candidate \mbox{\DzToKSKS} decays. The mass of the \Dz candidate, \mD, is required to be in the $[1.85,1.89]\gevcc$ range for Belle and in the $[1.85,1.88]\gevcc$ range for Belle II, corresponding to approximately three times the mass resolution. The different ranges account for a small offset observed in the \Dz mass-peak positions in the two datasets and for the different mass resolutions.

For the control mode decays, candidate \Dz mesons are formed by combining pairs of oppositely charged kaons with mass, \mKK, in the $[1.75,2.05]\gevcc$ range. Tracks originating from charged kaons must have at least 20 hits in the central drift chamber and at least one hit in the silicon vertex detector. They must have a distance of closest approach to the \epem interaction point (IP) smaller than 2.0\cm in the longitudinal direction (parallel to the solenoid axis and in the direction of the positron beam) and smaller than 0.5\cm in the transverse plane. We identify kaons by requiring $\mathcal{L}_{K}/(\mathcal{L}_{K} + \mathcal{L}_{\pi})>0.6$ and $\mathcal{L}_{K}/(\mathcal{L}_{K} + \mathcal{L}_{e})>0.1$, where $\mathcal{L}_x$ is the likelihood for the hypothesis $x$ to have produced the relevant track. The particle-identification likelihoods are based on information from the aerogel threshold Cherenkov counters, time-of-flight scintillation counters, and the central drift chamber for Belle; and from all subdetectors except the pixel detector for Belle~II. The kaon-identification efficiency is above 86\% with rates of pion-to-kaon and electron-to-kaon misidentification below 9\% and 20\%, respectively.

The \Dz candidates for both signal and control modes are then combined with low-momentum pions to form a $\Dstarp\to\Dz\pip$ decay. The soft-pion tracks must be in the acceptance of the central drift chamber and have longitudinal and transverse distances of closest approach to the IP smaller than 2.0\cm and 0.5\cm, respectively. To improve signal efficiency, particle identification is not required for the soft-pion candidates. The \Dstarp candidates undergo a kinematic fit~\cite{Krohn:2019dlq}, which constrains the \Dstarp vertex to the measured position of the IP, and the masses of the two \KS candidates to the nominal \KS mass~\cite{pdg} for the signal mode. Only candidates whose kinematic fits converge with $\chi^2$ probabilities larger than $10^{-3}$ are retained for further analysis.
 
The difference between the reconstructed \Dstarp and \Dz masses is required to be smaller than 0.16\gevcc. To suppress events in which the \Dstarp candidate comes from the decay of a \B meson, the momentum of the \Dstarp in the \epem center-of-mass system is required to to be greater than 2.5\gevc. Simulation shows that the remaining contamination from \B meson decays amounts to less than one candidate per 1\invab of integrated luminosity and is negligible.

\section{Kinematic weighting\label{sec:equalization}}
\begin{figure*}[t]
\centering
\includegraphics[width=.8\linewidth]{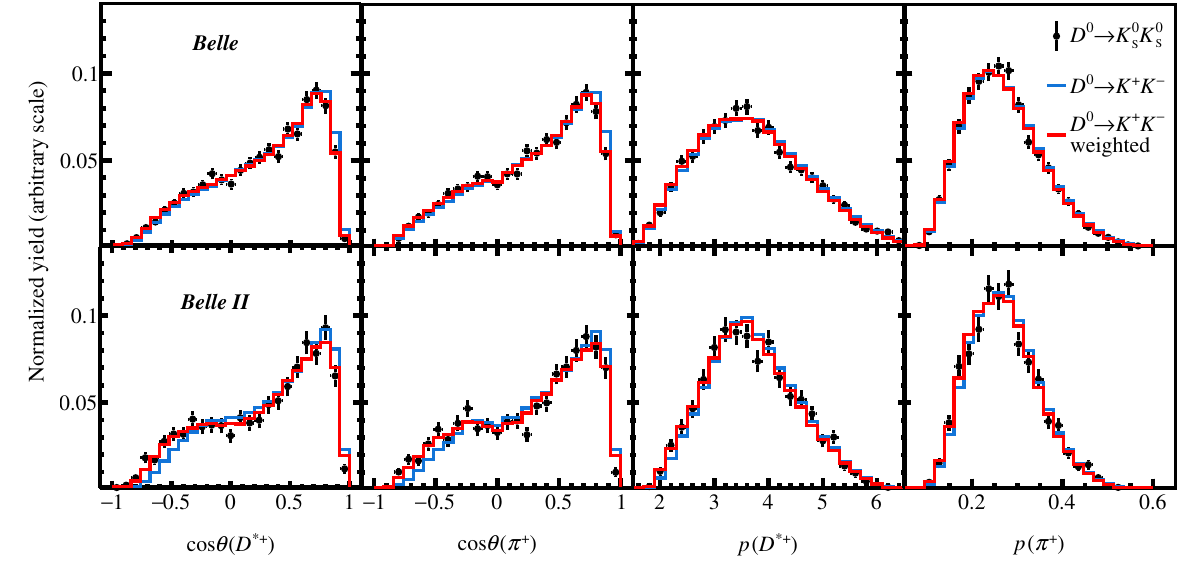}\\
\caption{Normalized distributions of (from left to right) cosine of the \Dstarp polar angle, cosine of the soft-pion polar angle, \Dstarp momentum, and soft-pion momentum of background-subtracted \DzToKSKS and \DzToKpKm decays in (top) Belle and (bottom) Belle II data, before and after the kinematic weighting.\label{fig:kinematic_comparison}}
\end{figure*}

Because detector- and production-induced asymmetries depend on kinematic properties of the selected \Dstarp candidates, the asymmetry cancellation is realized accurately only if the kinematic distributions in the \DzToKSKS and \DzToKpKm samples are the same. The \Dstarp production asymmetry is expected to vary as a function of the \Dstarp polar angle, $\theta(\Dstarp)$; the soft-pion detection asymmetry is expected to vary both as a function of momentum and of polar angle. Because of the differences in the reconstruction and selection of charged and neutral kaons, small differences are present in these distributions for the selected \DzToKSKS and \DzToKpKm candidates (\cref{fig:kinematic_comparison}). A weighting procedure is therefore implemented to reduce the observed differences. The ratio of the $\cos\theta(\Dstarp)$ distributions of \DzToKSKS and \DzToKpKm decays in simulation is used to determine a smooth curve that provides a candidate-specific weight for the \DzToKpKm sample in data. Weighting as a function of $\cos\theta(\Dstarp)$ also reduces the differences in the other kinematic distributions (\cref{fig:kinematic_comparison}). The \DzToKpKm distributions shown in the remainder of this paper are weighted according to this procedure.

\section{Determination of observed asymmetries\label{sec:fit}}
The asymmetries between the observed yields of \Dstarp and \Dstarm signal candidates are determined using unbinned maximum-likelihood fits to distributions that distinguish signal and control mode decays from background processes. The fit models are assumed to be the same for charm and anti-charm mesons. Their functional forms are extracted from either simulation, with parameter values that are adjusted for the data, or from sideband data.

\subsection{\boldmath Fit to the \DzToKSKS sample}
There are three components in the fit: signal \DzToKSKS decays, peaking background from \DzToKSpipi decays, and non-peaking background. Non-peaking background is separated from the other two components using the distribution of the \Dstarp mass, \mDzpi, which is calculated using the vector sum of the momenta of the three final-state particles as \Dstarp momentum, and the known \Dz mass~\cite{pdg} in the determination of the \Dstarp energy~\cite{CDF:2011ejf,DiCanto:2012upq}. Signal and peaking-background \DzToKSpipi decays have identical \mDzpi distributions, which narrowly peak at the nominal \Dstarp mass. In contrast, the non-peaking background has a smooth \mDzpi distribution. To separate signal from peaking-background, we use the large distance ($L$) between the \KS and \Dz decay vertices resulting from the long \KS lifetime. We introduce the variable $\Lminsig=\log\!\left[\min\!\left(L_1/\sigma_{L_1}, L_2/\sigma_{L_2} \right) \right]$, where $L_{1(2)}$ and $\sigma_{L_{1(2)}}$ are the distance and its uncertainty for the first (second) \KS candidate, respectively. Both the signal and peaking-background \Lminsig distributions exhibit a peaking structure, but they peak at very different values.

We determine the signal yield and raw asymmetry by fitting to the \mDzpi and \Lminsig distributions, simultaneously for \Dstarp and \Dstarm candidates. In the fit, the two-dimensional probability density functions (PDFs) of each component can be factorized into the product of one-dimensional PDFs, as verified in simulation. The \mDzpi and \Lminsig PDFs of signal and peaking-background decays are each modeled using Johnson's $S_{U}$ functions with parameters derived from simulation~\cite{johnson}. The \mDzpi PDF is assumed to be the same for \DzToKSKS and \DzToKSpipi decays; the \Lminsig PDF peaks at larger values for \DzToKSKS than for \DzToKSpipi decays. The \mDzpi distribution of the non-peaking background is modeled as a threshold-like distribution,
\begin{equation}\label{eq:threshold-function}
\pdf^\bkg(m) \propto (m-m_0)^{1/2}+\alpha(m-m_0)^{3/2}\,
\end{equation}
with the threshold parameter $m_0$ fixed to the known value corresponding to the sum of the nominal \Dz and charged-pion masses~\cite{pdg}. The parameter $\alpha$ is determined directly from the fit to the data, together with the yields and asymmetries of each component. The \Lminsig PDF is modeled as the sum of two Johnson's $S_U$ functions peaking at different values. All the parameter values are determined using candidates populating the \mDzpi sideband $[2.005, 2.008]\cup[2.013, 2.023]\gevcc$. Simulation shows that candidates in this sideband reproduce the distribution of the non-peaking background in the signal region.

\begin{figure*}[ht]
\centering
\includegraphics[width=.7\linewidth]{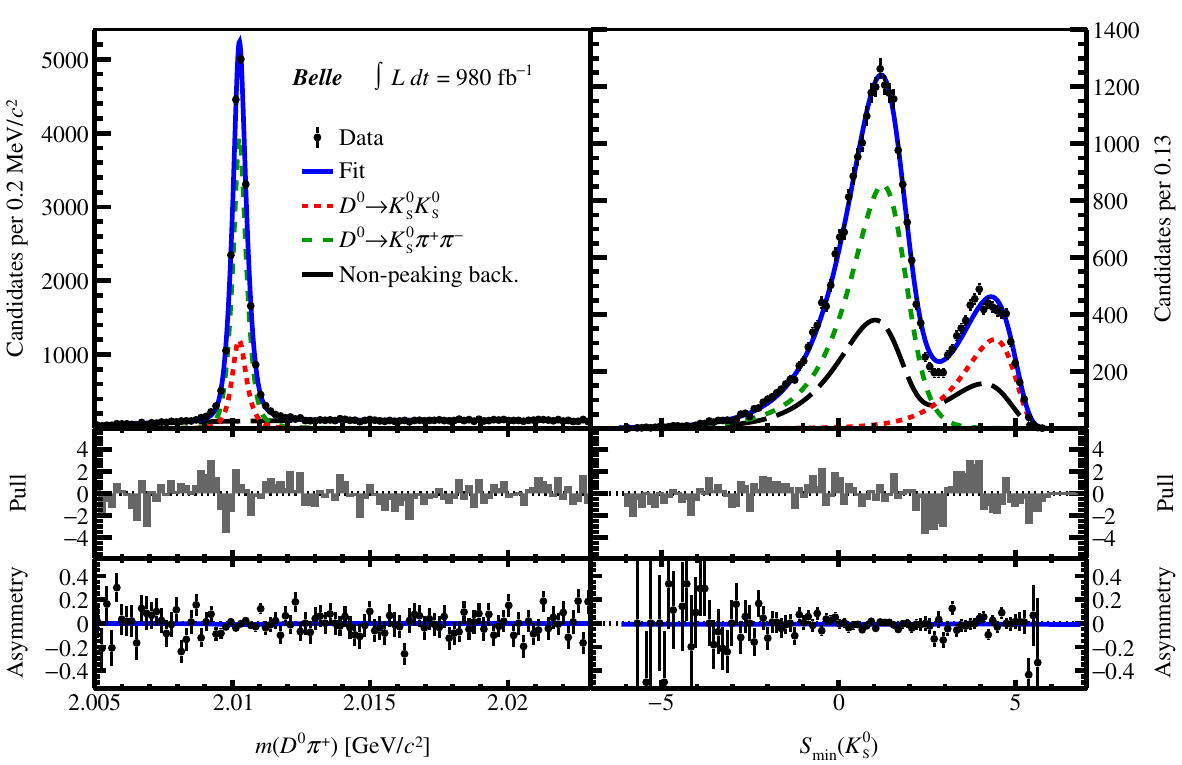}\hfil
\includegraphics[width=.7\linewidth]{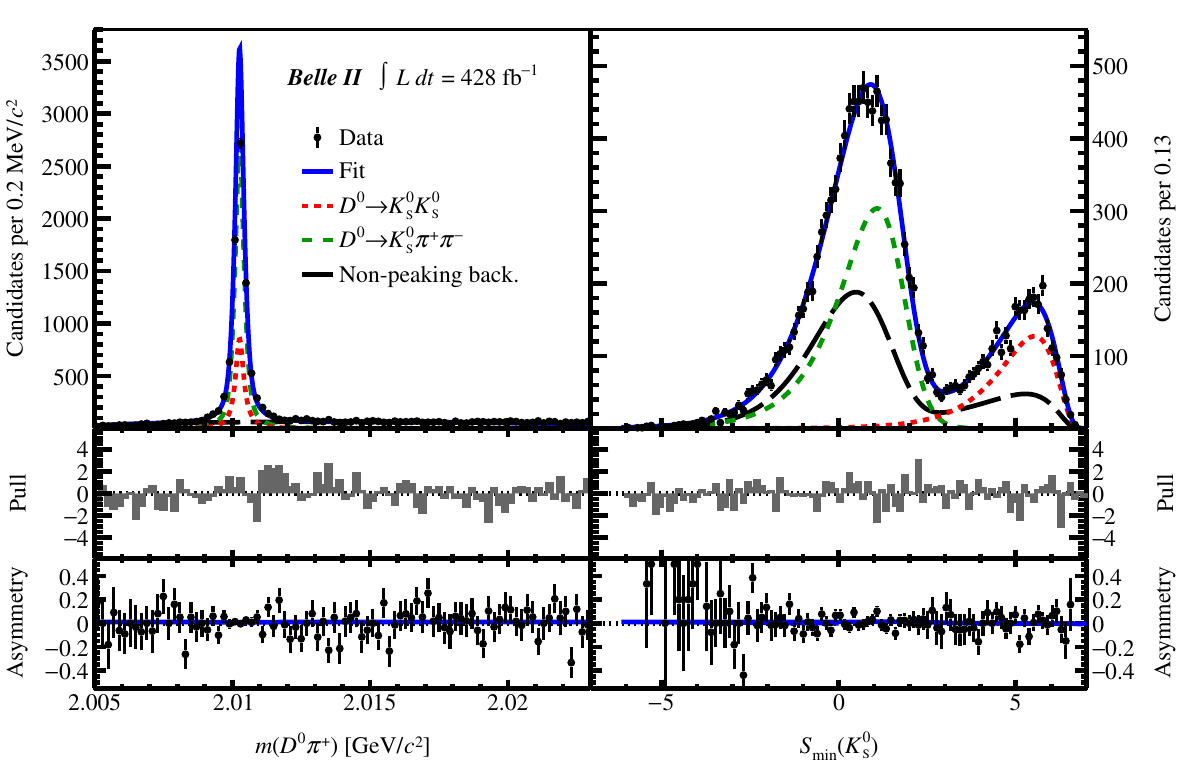}\\
\caption{Distributions of (left) \mDzpi and (right) \Lminsig for combined \DzToKSKS and \DzbToKSKS candidates, in (top) Belle and (bottom) Belle II data, with fit projections overlaid.  The middle panel of each plot shows the distribution of the difference between observed and fit values divided by the uncertainty from the fit (pull), the bottom panel shows the asymmetry between \Dz and \Dzb candidates with the fit projection overlaid. \label{fig:signal_fit}}
\end{figure*}

The results of the fit to the Belle and Belle II data are shown in \cref{fig:signal_fit}. The measured signal yields are $4864\pm78$ in Belle and $2214\pm51$ in Belle II. The raw asymmetry is measured to be $(-1.0\pm1.6)\%$ in Belle and $(-0.6\pm2.3)\%$ in Belle II. The uncertainties are statistical only. The fit model describes the data well, except for the Belle case in the region around $\Lminsig=3.5$. The mismodeling is similar for \Dstarp and \Dstarm candidates and hence does not significantly affect the asymmetry. Fits to simulation, which have a similar mismodeling, show no evidence of a bias in the determinations of the signal yield and asymmetry. A systematic uncertainty is assigned by repeating the fit to the data with alternative models, as discussed in \cref{sec:systematics}.

\afterpage{\clearpage}
\subsection{\boldmath Fit to the \DzToKpKm sample}
The raw asymmetry of the control decays is determined using a fit to the two-dimensional distribution of \mKK and \mDzpi. The fit consists of the following components: $\Dz\to\Kp\Km$ control mode decays; $\Dz\to\Km\pip$ decays in which the pion is misidentified as a kaon, peaking at \mKK values around 1.94\gevcc; partially reconstructed \mbox{$\Dz\to\text{multibody}$} decays, where the term multibody refers to decays such as $\Dz\to\Km\pip\piz$, where the charged pion is mis-identified as a kaon and the neutral pion is not reconstructed, or semileptonic \Dz decays, where the neutrino is not reconstructed, populating mostly the low-\mKK region; $\Dsp\to\Kp\Km\pip$ decays in which the \pip is reconstructed as a soft-pion candidate, peaking at \mKK values around 1.82\gevcc; and combinatorial background populating the entire \mKK range nearly uniformly. Each \Dz component can either be associated with a real soft-pion candidate, which peaks in \mDzpi, or with an unrelated soft-pion candidate, which contributes to a smoothly distributed random-pion background in \mDzpi. The \Dsp background has a distribution that rises almost linearly in \mDzpi. Simulation shows that the two-dimensional PDF of each component, except for the $\Dsp\to\Kp\Km\pip$ background, can be approximated by the product of two one-dimensional PDFs.

The control decays and physics backgrounds PDFs are determined using simulation. The \mKK and \mDzpi PDFs of the $\Dstarp\to\Dz(\to\Kp\Km)\pip$ decays are each modeled using the sum of two Gaussian functions (with a common mean) and of a Johnson's $S_{U}$ function. The \mKK PDF of the mis-identified $\Dstarp\to\Dz(\to\Km\pip)\pip$ background is parametrized using the sum of a Gaussian and a Johnson's $S_{U}$ function. Given that \mDzpi is unaffected by the misidentification of the \Dz final-state particles, the \mDzpi PDF is shared with that of the control decays. To account for mismodeling of the simulation, shape parameters related to the peak positions and resolutions of the $\Dz\to\Kp\Km$ and $\Dz\to\Km\pip$ components are floated when fitting to the data, while other parameters are fixed to the values obtained from simulation.

The partially reconstructed $\Dstarp\to\Dz(\to\text{multibody})\pip$ decays are modeled as an exponential function in \mKK and as a Johnson's $S_{U}$ function in \mDzpi, with parameters fixed to simulation.

For each of the aforementioned components there is a background in which an unrelated soft pion is associated with the identified \Dz candidate. These random-pion components share the same \mKK distribution as the component with the correctly reconstructed soft pion. Their \mDzpi distributions are modeled by the following common PDF:
\begin{equation}\label{eq:RooDstD0Bkg}
\pdf^{\text{rnd}}(m|m_0,A,B) \propto \left[1-\exp\!\left(-\frac{m-m_{0}}{A}\right)\right]\left(\frac{m}{m_{0}}\right)^{B}\,,
\end{equation}
with the parameters $A$ and $B$ free to float in the fit.

The $\Dsp\to\Kp\Km\pip$ background, in which the pion is used as the soft-pion candidate, exhibits a kinematic correlation between the average values of \mKK and \mDzpi which can be calculated analytically as $\mean{\mKK}(\mDzpi) = m_{\Dsp} + m_{\Dz} - \mDzpi$, using the known \Dsp and \Dz masses~\cite{pdg}. The two-dimensional PDF is written as the product of the \mKK PDF, conditional on the value of \mDzpi, and the \mDzpi PDF,
\begin{multline}
\pdf^{\Dsp}(\mKK,\mDzpi) =\\ \pdf^{\Dsp}(\mKK|\mDzpi)\, \pdf^{\Dsp}(\mDzpi)\,.
\end{multline}
The first term on the right-hand side is parametrized as a Johnson's $S_{U}$ function with a mean given by $\mu_J+\mean{\mKK}(\mDzpi)$, where $\mu_J$ is an offset which accounts for possible data-simulation differences in the peak position and is floated in the fit to the data. The \mDzpi PDF is a first-order polynomial defined only above the threshold value of $m_0$.

Finally, the combinatorial background PDF is modeled by a product of a linear function in \mKK and \cref{eq:RooDstD0Bkg} for \mDzpi. The parameters of the combinatorial background are floated in the fit to the data. The yield and asymmetry of each component are the remaining free parameters of the fit.

\begin{figure*}[ht]
\centering
\includegraphics[width=.7\linewidth]{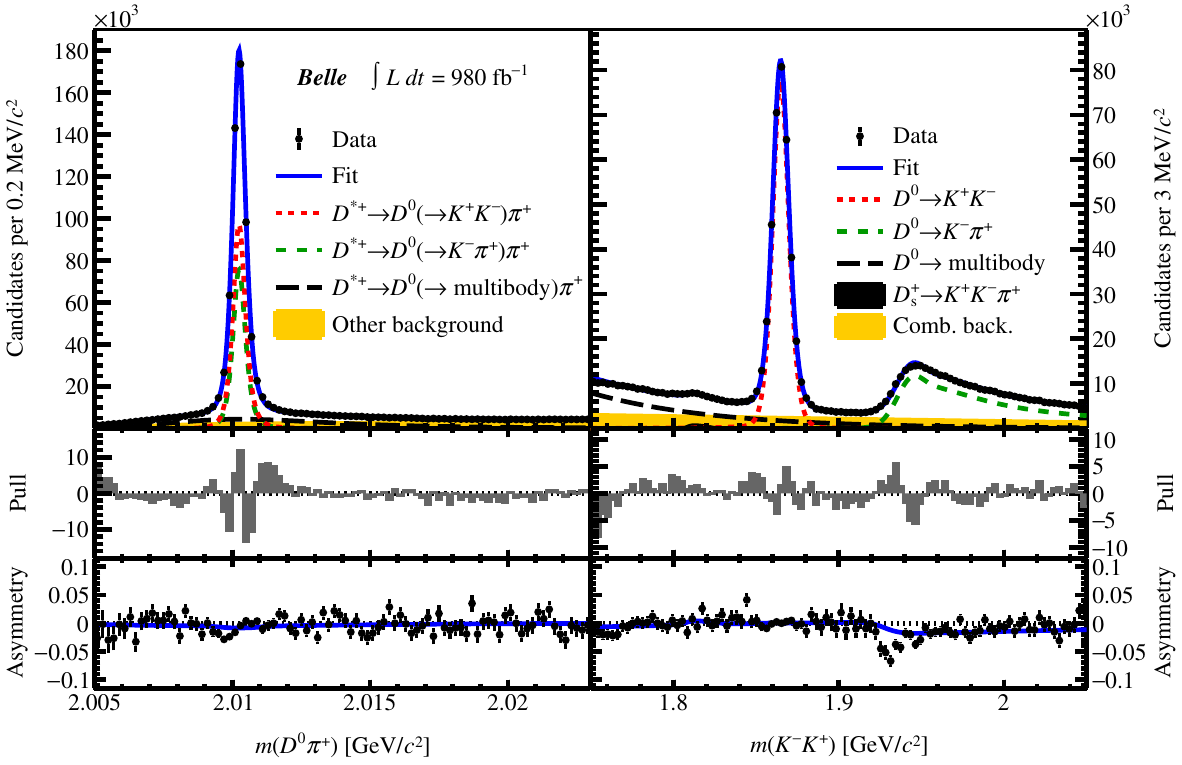}\hfil
\includegraphics[width=.7\linewidth]{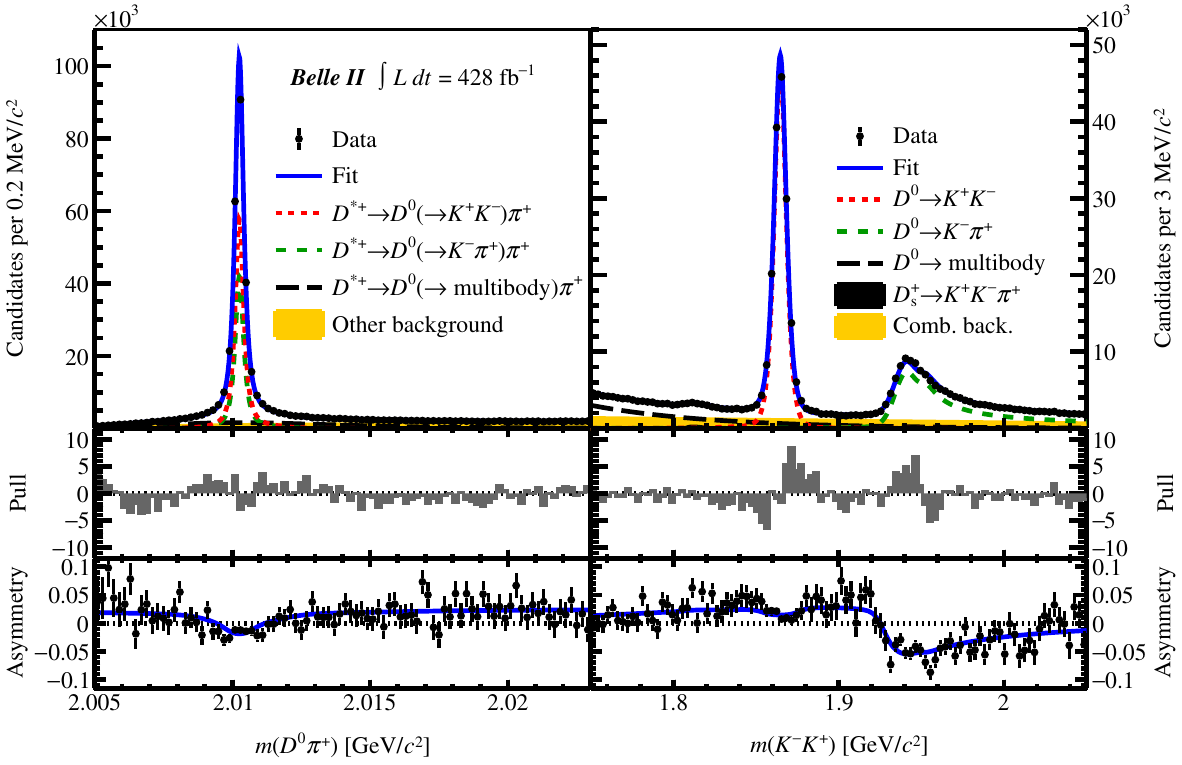}\\
\caption{Distributions of (left) \mDzpi and (right) \mKK for combined \DzToKpKm and \DzbToKpKm candidates, in (top) Belle and (bottom) Belle II data, with fit projections overlaid. The middle panel of each plot shows the distribution of the difference between observed and fit values divided by the uncertainty from the fit (pull), the bottom panel shows the asymmetry between \Dz and \Dzb candidates with the fit projection overlaid. Other background includes the $\Dsp\to\Kp\Km\pip$, random-pion, and combinatorial components.\label{fig:control_fit}}
\end{figure*}

The results of the fit to the Belle and Belle II data are shown in \cref{fig:control_fit}. The measured \KK yields are $308760\pm570$ in Belle and $145520 \pm 400$ in Belle II. The \KK raw asymmetry is measured to be $(0.17\pm0.19)\%$ in Belle and $(1.61\pm0.27)\%$ in Belle II. The uncertainties are statistical only. The raw-asymmetry values for Belle and Belle II are consistent with expected differences in reconstruction asymmetries for low-momentum pions between the two experiments. While the variation of the asymmetry as a function of the fitted observables is fairly well described by the fit model, the \CP-averaged distributions of the fitted observables are not. A similar mismodeling is observed in simulation and results in 0.1-2\% biases in the measured \KK yields. The mismodeling has a small effect on the measured asymmetries, as observed in simulation and quantified by refitting to the data with alternative models (\cref{sec:systematics}).

%\afterpage{\clearpage}
\section{Systematic uncertainties\label{sec:systematics}}
The following sources of systematic uncertainties are considered: PDF modeling in the \DzToKSKS and \DzToKpKm fits, kinematic weighting of the control and signal sample for the cancellation of the nuisance asymmetries, uncertainties on the external input value of $\Acp(\DzToKpKm)$. A summary of the estimated uncertainties is given in \cref{tab:syst}, the total is the sum in quadrature of the individual components.

\begin{table}[ht]
\centering
\caption{Summary of uncertainties in $\Acp(\DzToKSKS)$.\label{tab:syst}}
\begin{tabular}{lrr}
\hline\hline
Source & \multicolumn{2}{c}{Uncertainty (\%)} \\
 & Belle & Belle II\\
\hline
Modeling in the \DzToKSKS fit & 0.04 & 0.05 \\
Modeling in the \DzToKpKm fit & 0.02 & $<0.01$ \\
Kinematic weighting           & 0.06 & 0.07 \\
Input $\Acp(\DzToKpKm)$       & 0.05 & 0.05 \\
\hline
Total systematic              & 0.09 & 0.10 \\
Statistical                   & 1.60 & 2.30 \\
\hline\hline
\end{tabular}
\end{table}

To estimate the systematic uncertainty due to the fit models, we repeat the fit to the data using several variations of the default models. We test two classes of variations: one in which we vary the PDF model, which is the same for \Dz and \Dzb candidates, and another in which we keep the default PDF model but introduce different shape parameters for \Dz and \Dzb candidates. For the first class of variations we use (six for \DzToKSKS, and seven for \DzToKpKm) alternative models featuring similar or much worse description of the data compared to the default model. The alternative models change the $-2\log\mathcal{L}$ value, where $\mathcal{L}$ is the likelihood at the minimum, by an amount ranging between $-6$ and $86$ for the \KSKS fits and between $-377\times10^3$ and $35.9\times10^3$ for the \KK fits, while having similar numbers of free parameters as the nominal model. The largest variations in the measured raw asymmetries are assigned as systematic uncertainties. From the second class of model variations we observe no significant changes in the asymmetries and consistent fit qualities for alternative and default models. Hence, we assign no additional systematic uncertainty due to the default-fit assumption that \Dz and \Dzb shapes are the same.

The kinematic weighting of the \DzToKpKm control sample does not entirely remove all differences observed in the \Dstarp and soft-pion momentum and polar angle distributions. The residual differences are approximately a factor two smaller than those observed before the weighting. We therefore take half the variation in $\Araw^{\KK}$ introduced by the weighting procedure, 0.06\% in Belle and 0.07\% in Belle II, as a systematic uncertainty.

The uncertainty in the input value of $\Acp(\DzToKpKm)$, 0.05\%~(\cref{eq:inputAcpD0KK}), is propagated as a systematic uncertainty on the measurement.

The entire analysis procedure, including the kinematic weighting of the control mode, is validated using sets of pseudoexperiments generated by sampling from the fit PDFs and using fully simulated decays. In both cases, the results of the validation confirm that we estimate the \CP asymmetry in \DzToKSKS decays, and its uncertainty, without bias.

Finally, we check the consistency of the measured value of $\Acp(\DzToKSKS)$ by repeating the measurement in disjoint subsets of the data split according to data-taking conditions or momentum of the \Dstarp candidate. In all cases, observed variations in the results are consistent with each other and with the measurement from the full sample.

\section{Final results and conclusions\label{sec:results}}
Using \Dstarp-tagged \DzToKSKS and \DzToKpKm decays reconstructed in \BelleLumi of Belle data, and in \BelleIILumi of Belle II data, we measure the time-integrated \CP asymmetry in \DzToKSKS decays to be
\begin{align}
\Acp(\DzToKSKS) &= (\resBelle\pm\statBelle\stat\pm\systBelle\syst)\%
\intertext{and}
\Acp(\DzToKSKS) &= (\resBelleII\pm\statBelleII\stat\pm\systBelleII\syst)\%\,,
\end{align}
respectively. The two results are consistent and agree with previous determinations~\cite{CLEO:2000opx,LHCb:2015ope,Dash:2017heu,LHCb:2021rdn,CMS:2024hsv}. The result based on Belle data supersedes the published result~\cite{Dash:2017heu}. It has a factor of two smaller systematic uncertainty compared to the previous result thanks to the usage of the \DzToKpKm control mode, which provides a more precise \Acp external input compared to the $\Dz\to\KS\piz$ control mode used in Ref.~\cite{Dash:2017heu}.  Assuming that the only source of correlation between the two results is the external input value of $\Acp(\DzToKpKm)$, we combine our Belle and Belle II results to obtain
\begin{equation}
\Acp(\DzToKSKS) = (\resComb\pm\statComb\stat\pm\systComb\syst)\%\,.
\end{equation}
The combined result has precision comparable to the world's best measurement from LHCb~\cite{LHCb:2021rdn}, with which it agrees.

% Policy from October 20, 2022
This work, based on data collected using the Belle II detector, which was built and commissioned prior to March 2019,
and data collected using the Belle detector, which was operated until June 2010,
was supported by
%Armenia
Higher Education and Science Committee of the Republic of Armenia Grant No.~23LCG-1C011;
%Australia
Australian Research Council and Research Grants
No.~DP200101792, % Jackson
No.~DP210101900, % Urquijo
No.~DP210102831, % Sevior
No.~DE220100462, % Hsu
No.~LE210100098, % Infrastructure
and
No.~LE230100085; % Infrastructure
%Austria
Austrian Federal Ministry of Education, Science and Research,
Austrian Science Fund
No.~P~34529,
No.~J~4731,
No.~J~4625,
and
No.~M~3153,
and
Horizon 2020 ERC Starting Grant No.~947006 ``InterLeptons'';
%Canada
Natural Sciences and Engineering Research Council of Canada, Compute Canada and CANARIE;
%China
National Key R\&D Program of China under Contract No.~2022YFA1601903,
National Natural Science Foundation of China and Research Grants
No.~11575017,
No.~11761141009,
No.~11705209,
No.~11975076,
No.~12135005,
No.~12150004,
No.~12161141008,
No.~12475093,
and
No.~12175041,
and Shandong Provincial Natural Science Foundation Project~ZR2022JQ02;
%Czech Republic
the Czech Science Foundation Grant No.~22-18469S 
and
Charles University Grant Agency project No.~246122;
%EU
European Research Council, Seventh Framework PIEF-GA-2013-622527,
Horizon 2020 ERC-Advanced Grants No.~267104 and No.~884719,
Horizon 2020 ERC-Consolidator Grant No.~819127,
Horizon 2020 Marie Sklodowska-Curie Grant Agreement No.~700525 ``NIOBE''
and
No.~101026516,
and
Horizon 2020 Marie Sklodowska-Curie RISE project JENNIFER2 Grant Agreement No.~822070 (European grants);
%France
L'Institut National de Physique Nucl\'{e}aire et de Physique des Particules (IN2P3) du CNRS
and
L'Agence Nationale de la Recherche (ANR) under grant ANR-21-CE31-0009 (France);
%Germany
BMBF, DFG, HGF, MPG, and AvH Foundation (Germany);
%India
Department of Atomic Energy under Project Identification No.~RTI 4002,
Department of Science and Technology,
and
UPES SEED funding programs
No.~UPES/R\&D-SEED-INFRA/17052023/01 and
No.~UPES/R\&D-SOE/20062022/06 (India);
%Israel
Israel Science Foundation Grant No.~2476/17,
U.S.-Israel Binational Science Foundation Grant No.~2016113, and
Israel Ministry of Science Grant No.~3-16543;
%Italy
Istituto Nazionale di Fisica Nucleare and the Research Grants BELLE2;
%Japan
Japan Society for the Promotion of Science, Grant-in-Aid for Scientific Research Grants
No.~16H03968,
No.~16H03993,
No.~16H06492,
No.~16K05323,
No.~17H01133,
No.~17H05405,
No.~18K03621,
No.~18H03710,
No.~18H05226,
No.~19H00682, % Niigata
No.~20H05850,
No.~20H05858,
No.~22H00144,
No.~22K14056,
No.~22K21347,
No.~23H05433,
No.~26220706,
and
No.~26400255,
%the National Institute of Informatics, and Science Information NETwork 5 (SINET5), 
and
the Ministry of Education, Culture, Sports, Science, and Technology (MEXT) of Japan;  
%Korea
National Research Foundation (NRF) of Korea Grants
No.~2016R1-D1A1B-02012900,
No.~2018R1-A6A1A-06024970,
No.~2021R1-A6A1A-03043957,
No.~2021R1-F1A-1060423,
No.~2021R1-F1A-1064008,
No.~2022R1-A2C-1003993,
No.~2022R1-A2C-1092335,
No.~RS-2023-00208693,
No.~RS-2024-00354342
and
No.~RS-2022-00197659,
Radiation Science Research Institute,
Foreign Large-Size Research Facility Application Supporting project,
the Global Science Experimental Data Hub Center, the Korea Institute of
Science and Technology Information (K24L2M1C4)
and
KREONET/GLORIAD;
%Malaysia
Universiti Malaya RU grant, Akademi Sains Malaysia, and Ministry of Education Malaysia;
%Mexico
% CINVESTAV-IPN, UNAM, UAS, BUAP and CONACYT are funded under
Frontiers of Science Program Contracts
No.~FOINS-296,
No.~CB-221329,
No.~CB-236394,
No.~CB-254409,
and
No.~CB-180023, and SEP-CINVESTAV Research Grant No.~237 (Mexico);
%Poland
the Polish Ministry of Science and Higher Education and the National Science Center;
%Russia
the Ministry of Science and Higher Education of the Russian Federation
and
the HSE University Basic Research Program, Moscow;
%Saudi Arabia
University of Tabuk Research Grants
No.~S-0256-1438 and No.~S-0280-1439 (Saudi Arabia);
%Slovenia
Slovenian Research Agency and Research Grants
No.~J1-9124
and
No.~P1-0135;
%Spain
Ikerbasque, Basque Foundation for Science,
the State Agency for Research of the Spanish Ministry of Science and Innovation through Grant No. PID2022-136510NB-C33,
Agencia Estatal de Investigacion, Spain
Grant No.~RYC2020-029875-I
and
Generalitat Valenciana, Spain
Grant No.~CIDEGENT/2018/020;
%Swiss (Belle 1)
the Swiss National Science Foundation;
%Sweden
The Knut and Alice Wallenberg Foundation (Sweden), Contracts No.~2021.0174 and No.~2021.0299;
%Taiwan
National Science and Technology Council,
and
Ministry of Education (Taiwan);
%Thailand
Thailand Center of Excellence in Physics;
%Turkey
TUBITAK ULAKBIM (Turkey);
%Ukraine
National Research Foundation of Ukraine, Project No.~2020.02/0257,
and
Ministry of Education and Science of Ukraine;
%USA
the U.S. National Science Foundation and Research Grants
No.~PHY-1913789 % Indiana CEEM
and
No.~PHY-2111604, % Luther
and the U.S. Department of Energy and Research Awards
No.~DE-AC06-76RLO1830, % PNNL
No.~DE-SC0007983, % Wayne State
No.~DE-SC0009824, % Florida
No.~DE-SC0009973, % VPI
No.~DE-SC0010007, % Duke
No.~DE-SC0010073, % South Carolina
No.~DE-SC0010118, % Carnegie Mellon
No.~DE-SC0010504, % Hawaii
No.~DE-SC0011784, % Cincinnati
No.~DE-SC0012704, % BNL
No.~DE-SC0019230, % Duke
No.~DE-SC0021274, % Mississippi
No.~DE-SC0021616, % Mississippi
No.~DE-SC0022350, % Louisville
No.~DE-SC0023470; % South Alabama
%last group
and
%Vietnam
the Vietnam Academy of Science and Technology (VAST) under Grants
No.~NVCC.05.12/22-23
and
No.~DL0000.02/24-25.

% Policy from October 20, 2022
These acknowledgements are not to be interpreted as an endorsement of any statement made
by any of our institutes, funding agencies, governments, or their representatives.

We thank the SuperKEKB team for delivering high-luminosity collisions;
the KEK cryogenics group for the efficient operation of the detector solenoid magnet and IBBelle on site;
the KEK Computer Research Center for on-site computing support; the NII for SINET6 network support;
and the raw-data centers hosted by BNL, DESY, GridKa, IN2P3, INFN, 
PNNL/EMSL, 
and the University of Victoria.

\bibliographystyle{belle2}
\bibliography{references}

\end{document}